\documentclass[letterpaper,11pt,headings=standardclasses,parskip=half]{scrartcl}
\makeatletter\if@twocolumn
	\providecommand{\twocoldoc}{}
	
	\providecommand{\nogeometry}{}
	\providecommand{\nocustomize}{}
	\providecommand{\blind}{}
\fi\makeatother

\usepackage{amssymb}
\usepackage{authblk}
\usepackage{comment}
\usepackage{caption}
\usepackage{enumitem}
\usepackage{filecontents}
\usepackage{hyperref}
\ifdefined\nogeometry\else\usepackage[nohead, margin=1in]{geometry}\fi
\usepackage{graphicx}  
\usepackage[frozencache=true]{minted}  
\ifdefined\twocoldoc\else\usepackage{multicol}\fi
\ifdefined\blind\else\usepackage[numbers]{natbib}\fi
\usepackage{siunitx}
\usepackage{tabulary}
\ifdefined\nocustomize\else\usepackage{titling}\fi
\usepackage{url}
\usepackage{xcolor}

\ifdefined\KOMAClassName
	\setkomafont{title}{\normalfont\bfseries}
	\makeatletter
	\renewcommand{\maketitle}{
		\begin{center}
			\vspace*{\droptitle}
			\maketitlehooka
			{\Large\bfseries\@title\par\vspace*{0.5\baselineskip}}
			\maketitlehookb
			{\@author}
			\maketitlehookc
			{\@date}
			\maketitlehookd
		\end{center}
	}
	\makeatother
\fi

\definecolor{links}{HTML}{2A1B81}
\hypersetup{pdfpagemode=UseNone,pdfpagelayout=OneColumn,pdfstartview=,colorlinks,linkcolor=,urlcolor=links}
\newminted                    {cpp}{mathescape, numbersep=0.5em, escapeinside=??, linenos=false, numberblanklines=false, tabsize=4}
\newminted[python2indent4]{python }{mathescape, numbersep=0.5em, escapeinside=??, linenos=false, numberblanklines=false, tabsize=4, python3=false}
\newminted[python3indent4]{python3}{mathescape, numbersep=0.5em, escapeinside=??, linenos=false, numberblanklines=false, tabsize=4, python3=true}
\newminted[python3indent8]{python3}{mathescape, numbersep=0.5em, escapeinside=??, linenos=false, numberblanklines=false, tabsize=8, python3=true}
\setminted{fontsize=\small}

\ifdefined\nocustomize\else\fi

\ifdefined\nocustomize\else
	\AfterEndEnvironment{minted}{\vspace*{-1\baselineskip}}
\fi

\DeclareRobustCommand{\CS}{C\#}
\DeclareRobustCommand{\CCLarge}{C\nolinebreak[4]\hspace{-.00em}\raisebox{.2ex}{\normalsize ++}}
\DeclareRobustCommand{\CC     }{C\nolinebreak[4]\hspace{-.00em}\raisebox{.2ex}{\small      ++}}

\newenvironment{longlisting}{\captionsetup{type=listing,}}{}

\setlist[itemize]{noitemsep, topsep=1pt}

\title{\textbf{Inflationary Constant Factors\\And Why Python is Faster Than \CCLarge{}}}
\date{}
\ifdefined\blind\else
	\author{Mehrdad Niknami}
	\affil{University of California, Berkeley}
\fi

\begin{document}
	\maketitle

\begin{abstract}
	Constant-factor differences are frequently ignored when analyzing the complexity
	of algorithms and implementations, as they appear to be insignificant in practice.
	In this paper, we demonstrate that this assumption can in fact have far more profound
	implications on time complexity than is obvious at first glance,
	and that a poor consideration of trade-offs can result in \textit{polynomially}
	slower algorithms whose roots can be deeply and fundamentally ingrained into a
	programming language itself.
	While the general observation may not be novel from a theoretical standpoint, it is
	rarely (if ever) presented in traditional computer science curricula or other settings,
	and appears to be far from common knowledge in practical software engineering.
	We thus hope bring awareness to this issue and urge careful consideration of
	significant trade-offs that can result from trivial decisions made while programming.
\end{abstract}

\section{Introduction}
	There are often multiple competing algorithms for solving the same problem,
	with various trade-offs between them.
	In many cases, different methods for solving the same problem appear to have
	the same time complexity to within a constant factor.
	Often times, when the difference does not appear to be too large,
	it is then disregarded, and an algorithm is then selected based on other considerations,
	such as its space complexity or the simplicity of its implementation.
	Multi-pass algorithms in particular are frequently preferred over streaming or
	single-pass algorithms due to their simplicity,
	even in cases where they may be a few times slower.

	In this paper, we show how the constant-factor nature
	of a slowdown may be deceptive, and demonstrate that it can result in unexpectedly
	\emph{polynomial} slowdown of an algorithm.
	While this is an insignificant observation from the standpoint of algorithmic analysis,
	we hope to raise awareness of the need to carefully scrutinize constant-factor differences
	that are traditionally disregarded in both the teaching of theoretical computer science
	and the course of practical software engineering,
	as their transformation into polynomial slowdowns does not appear to be common knowledge,
	but can have significant adverse usability---and even security---implications.

\section{Comparisons}

	We illustrate the problem by considering the case of comparison operators.
	Many data structures and algorithms, such as binary search trees and mergesort,
	require a partial or total order to be defined on the objects of interest.
	The objects of interest may be further composed of other objects,
	and thus may in turn require the definition of an ordering among those objects themselves.
	It frequently becomes expedient, therefore, to define a simple interface for comparisons,
	and to expect and implement this interface across all data types of interest.

\subsection{Types of Comparisons}

	Broadly, there are two popular competing interfaces for comparison operators:
	\begin{itemize}[noitemsep]
		\item 2-way comparisons (e.g. returning \verb|True| iff one object is less-than another, or \verb|False| otherwise)
		\item 3-way comparisons (returning $-1$ for less-than, $+1$ for greater-than, and $0$ otherwise)
	\end{itemize}
	Different languages and libraries prefer different approaches, and sometimes change their
	preferences over time to one or the other.
	For example, in Python~3 and \CC17 one implements 2-way comparison operators (e.g. \verb|__lt__| or \verb|<|)
	whereas in Python~2, \CS, and \CC20 one implements 3-way operators (e.g. \verb|CompareTo|, \verb|cmp|, or \verb|<=>|).
	Reasons cited for doing so include simplicity (for the former~\cite{python3.0:comparison}),
	and efficiency (for the latter~\cite{sutter:comparison}).

	Consider, then, how one may implement each of these interfaces in terms of the other.
	A typical implementation of lexicographical comparison
	(such as \CC's \verb|std::|\allowbreak{}\verb|lexicographical_compare|) may be similar to the following
	Python code. We are interested in its \textit{worst-case time complexity}.
\begin{longlisting}
	\ifdefined\twocoldoc\else\begin{multicols}{2}\fi
		\begin{python3indent4}
def cmp2(a, b):
	return lt2(b, a) - lt2(a, b)
# Uses 2-way < for primitives
def lt2(a, b):
	if not isinstance(a, list):
		global c; c += 1
		return a < b  # built-in
	for x, y in zip(a, b):
		if lt2(x, y): return True
		if lt2(y, x): break
	return False
		\end{python3indent4}
	\ifdefined\twocoldoc\else\vfill\null\columnbreak\fi
		\begin{python3indent4}
def lt3(a, b):
	return cmp3(a, b) < 0
# Uses 3-way cmp() for primitives
def cmp3(a, b):
	if not isinstance(a, list):
		global c; c += 1
		return cmp(a, b)
	for x, y in zip(a, b):
		r = cmp3(x, y)
		if r != 0: return r
	return 0
		\end{python3indent4}
	\ifdefined\twocoldoc\else\end{multicols}\fi
\end{longlisting}

\subsection{Analysis}

	At a cursory glance, it may appear as though \verb|lt2| and \verb|lt3|
	are approximately equally efficient.
	And indeed, the need for any analysis is often not apparent at this stage,
	and thus one of the two approaches is selected based on secondary considerations or
	preferences.

	However, a simple example illustrates that
	the use of 2-way comparisons can result in making \textit{two} passes rather
	than \textit{one} pass over an equal prefix:
	\ifdefined\twocoldoc\else\begin{multicols}{2}\fi
	\begin{python3indent4}
# 2 passes over common "a" prefix
lt2(["a", "b"], ["a", "c"])
# 1 pass   over common "a" prefix
lt3(["a", "b"], ["a", "c"])
	\end{python3indent4}
	\ifdefined\twocoldoc\else\end{multicols}\fi
	Thus \verb|lt2| is a multi-pass algorithm, whereas \verb|lt3| is a single-pass algorithm.
	Often the difference is negligible for small amounts of data.
	However, in situations where the amount of data to be processed is large---or where
	comparison operations are expensive for other reasons---this can result in a significant
	performance overhead, both in terms of running time (e.g. data may no longer fit in cache)
	and in terms of memory (e.g. data may no longer be streamed).
	The potential overhead of making multiple passes has been one of the motivations
	for 3-way comparisons in \CC20~\cite{sutter:comparison}.

\subsection{Unexpected Implications}

	Unfortunately, this appears to be the point at which analysis of the trade-offs is halted.
	When the performance overhead is noted, it is generally presented as a constant-factor
	issue, with no hint of possibly more drastic implications for running time.
	However, crucially, this fails to capture the full repercussions of using a multi-pass
	algorithm.
	To see why, it is helpful to think \textit{adversarially}.
	How might a malicious attacker exploit the constant-factor difference to make 2-way
	comparisons \textit{drastically} slower than 3-way comparisons?
	Well, one way to do so would be to exploit a hidden assumption:
	that the data structure is \textit{shallow} (and, often, of constant depth).

	To that end, consider what happens when we construct a deep \textit{tree}.
\begin{longlisting}
	\ifdefined\twocoldoc\else\begin{multicols}{2}\fi
		\begin{python3indent4}
def test(obj):
	r = []
	for lt in [lt2, lt3]:
		global c; c = 0
		t = default_timer()
		lt(obj, obj)
		t = default_timer() - t
		r += [c, t * 1E6]
	return tuple(r)
		\end{python3indent4}
	\ifdefined\twocoldoc\else\vfill\null\columnbreak\fi
		\begin{python3indent4}
from timeit import default_timer

def tree(h):
	if h == 0: return 0
	return [tree(h - 1), tree(h - 1)]

for h in range(6):
	print("%4d\t%4.0f us | %3d\t%2.0f us"
	      % test(tree(h)))
		\end{python3indent4}
	\ifdefined\twocoldoc\else\end{multicols}\fi
\end{longlisting}
	\begin{figure}[H]
		\centering
		\begin{tabulary}{1\columnwidth}{|R|R|R|R|R|R|} \hline
	Height ($h$) & Nodes ($n$) & \texttt{lt2} calls  & \texttt{lt2} time & \texttt{lt3} calls & \texttt{lt3} time \\\hline
	$0$ & $ 1$ & $   1$ &  $\SI{  2}{\micro\second}$ &  $ 1$ & $\SI{ 2}{\micro\second}$ \\\hline
	$1$ & $ 3$ & $   4$ &  $\SI{  3}{\micro\second}$ &  $ 2$ & $\SI{ 2}{\micro\second}$ \\\hline
	$2$ & $ 7$ & $  16$ &  $\SI{  7}{\micro\second}$ &  $ 4$ & $\SI{ 3}{\micro\second}$ \\\hline
	$3$ & $15$ & $  64$ &  $\SI{ 27}{\micro\second}$ &  $ 8$ & $\SI{ 7}{\micro\second}$ \\\hline
	$4$ & $31$ & $ 256$ &  $\SI{108}{\micro\second}$ &  $16$ & $\SI{13}{\micro\second}$ \\\hline
	$5$ & $63$ & $1024$ &  $\SI{429}{\micro\second}$ &  $32$ & $\SI{26}{\micro\second}$ \\\hline
		\end{tabulary}
	\end{figure}
	\noindent The worst-case number of comparisons performed for a binary tree of $n$ nodes ($n = 2^{h + 1} - 1$) is:
	\begin{itemize}
		\item About $n^2$ comparisons for 2-way comparisons
		\item About $n  $ comparisons for 3-way comparisons
	\end{itemize}
	Suddenly, the choice between 2-way comparisons and 3-way comparisons is
	now far less subjective than it may have first appeared:
	one becomes a quadratic-time algorithm, while the other stays a linear-time one.\footnote
	{
		Also notice that, if we deduplicate the nodes and turn the tree into a DAG,
		this becomes a reduction from an approximately $4^n$-time algorithm
		to a $2^n$-time one.
	}
	Further analysis reveals that, in a more general setting of a tree of $n$ nodes
	with branching factor $b$, the number of 2-way comparisons is about $n^{1+\log_2 b}$.

	Of particular noteworthiness is the fact that it is extremely easy for an algorithm to
	degrade into quadratic-like behavior \textit{gradually}---i.e., even if the absolute
	worst-case scenario of completely equal trees is unexpected in a particular application
	(although in other applications, it can in fact be quite common),
	similar scenarios sharing \textit{some} equals elements---even for trees of bounded
	depth---would incur similar quadratic-like behavior at those cases, dramatically
	degrading performance.
	Therefore, a rarity of the most pathological case does not mean that the larger
	problem is necessarily unlikely to occur in practice.

\subsection[Python vs. C++ Design and Performance]{Python vs. \CC{} Design and Performance}

	We can now easily see how the precise choice of comparison primitives can unexpectedly
	harm asymptotic performance:
	this can occur when comparing object graphs of large depth, such as large trees or DAGs.
	As an interesting case study, we analyze how the ordering and comparison models in \CC{}
	and Python affect the performance of lexicographical ordering algorithms built into each
	language.

\subsubsection[Comparisons in C++]{Comparisons in \CC}

	The design of the \CC{} comparisons has been quite simple until \CC17: classes have
	merely had to define a $<$ operator, from which other operations are logically deduced,
	resulting in a similar kind of behavior as in \verb|lt2|.

\subsubsection{Comparisons in Python}

	The design of Python has been somewhat different, and has evolved over time.
	Python~2 was originally designed around 3-way comparison operators,
	defined via the \verb|__cmp__| method~\cite{python2.0:comparison}.
	However, 2-way ``rich comparison'' methods (\verb|__lt__|, \verb|__eq__|)
	were soon introduced in Python~2.1,
	to address shortcomings related to numerical Python~\cite{python2.1:comparison}.
	Python~3 later deprecated \verb|__cmp__| in favor of
	2-way comparisons~\cite{python3.0:comparison}.

	Internally, Python~2 and Python~3 both take a slightly different
	approach to the implementation of lexicographical ordering than \verb|lt2| does:
	when implementing an operation such as \verb|list.__lt__|,
	rather than relying solely on a single 3-way or 2-way comparison operator,
	Python invokes \textit{both} 2-way inequality \textit{and} equality operators.
	Specifically, instead of using $\nless$ to deduce $\ngtr$ and $=$,
	it uses $\nless$ and $=$ to deduce $\ngtr$.
	The rationale for this appears to have been to allow for more efficient equality comparisons
	than ordering comparisons in certain scenarios, such as for short
	circuiting cases where the lengths of containers are unequal~\cite{pep207}.
	Indeed, this has had a somewhat unintuitive side-effect: compared to \verb|lt2|,
	the implementation is faster if \verb|__le__| and \verb|__eq__| are implemented directly,
	but slower if only \verb|__cmp__| is available.

	It appears, therefore, that the asymptotic benefit in the worst-case scenario
	may have been a happy accident resulting from
	a desire to improve empirical performance in certain classes of typical cases,
	although it is difficult to know this for certain.
	Whether deliberate or not, though, the gain in performance is quite real, and
	has resulted in a rather bizarre situation where one can seriously claim that
	Python can be thousands of times (even arbitrarily many times)
	\textit{faster} than a traditionally ``fast language'' like \CC---merely due to one of
	the fundamental decisions made either intentionally or unintentionally in its core design.

	A comparison of the performance of lexicographical ordering algorithms in \CC{} and
	Python vividly illustrates this fact.
\begin{longlisting}
	\ifdefined\twocoldoc\else\begin{multicols}{2}\fi
		\begin{python2indent4}
# Python version
from timeit import default_timer
class Tree(list):
	pass
def tree(h):
	s = Tree()
	if h > 0:
		s += [tree(h - 1), tree(h - 1)]
	return s
def main():
	h = 15
	s1 = tree(h)
	s2 = tree(h)
	t = default_timer()
	s1 < s2
	d = default_timer() - t
	print(d)                      # 0.0018s
main()
		\end{python2indent4}
	\ifdefined\twocoldoc\else\end{multicols}\fi
	\ifdefined\twocoldoc\else\begin{multicols}{2}\fi
		\begin{cppcode}
// C++ version
#include <time.h>
#include <iostream>
#include <vector>
struct Tree : std::vector<Tree> { };
Tree tree(size_t h)
{
	Tree s;
	if (h > 0)
		s.insert(s.end(), 2, tree(h - 1));
	return s;
}
		\end{cppcode}
		\begin{cppcode}
?\phantom{.}?
int main()
{
	size_t h = 15;
	Tree s1 = tree(h);
	Tree s2 = tree(h);
	clock_t t = clock();
	s1 < s2;
	double d = clock() - t;
	d /= CLOCKS_PER_SEC;
	std::cout << d << std::endl;  // 4.1s
}
		\end{cppcode}
	\ifdefined\twocoldoc\else\end{multicols}\fi
\end{longlisting}
	This algorithm takes increasingly longer to execute in \CC17 than in Python as
	$h$ increases.

\subsubsection{Contrasting Comparisons}

	Is the design of Python comparisons therefore strictly superior to that of \CC?
	Not quite. Aside from the (minor) inconvenience of defining multiple comparison operators,
	the decision to use \verb|__eq__| when defining ordering relations is based
	on a hidden constraint that is frequently \textit{semantically incorrect}:
	namely, on the tacit assumption that two unordered objects are in fact always equal.
	This is frequently not the case (for example, consider case-insensitive string comparisons),
	and it can be necessary to impose an ordering on objects based on completely different
	criteria than semantic equality (for example, consider integers ordered by their parity).
	Such a distinction is one motivation for the inclusion of partial and total orders
	as first-class citizens in the \CC20 standard.

	From the above discussion, We can see that there can be significant trade-offs regarding
	semantic and algorithmic complexity in the design and implementation of even the most
	basic and fundamental operations in computer science.
	These considerations need to be carefully accounted for in order to avoid
	permanently saddling a language or an implementation with potentially
	subtle and adverse performance or usability consequences.

\section{Hashing}
	While lexicographical comparison illustrates the problem quite nicely, it is far from the
	only setting in which constant-factor performance differences may be magnified into
	polynomial ones.
	To illustrate that this is not an isolated problem, we present another example,
	this time based on hashing.

\subsection{Counting Object Occurrences}
	Consider the case of counting the occurrences of items in a sequence.
\begin{longlisting}
	\begin{python3indent4}
def count(items):
	seen = {}  # hashtable
	for item in items:
		seen[item] = seen.get(item, 0) + 1
	return seen
	\end{python3indent4}
\end{longlisting}\ifdefined\nocustomize\else\vspace*{1\baselineskip}\fi
	Here, \verb|item| is hashed twice---once to retrieve its value,
	and once to increment the value associated with that key.
	We seek to exploit this fact to incur a polynomial slowdown.

\subsection{Analysis}
	Consider what happens if \verb|item| is \textit{itself} a sequence whose hash function
	uses this function:
\begin{longlisting}
	\ifdefined\twocoldoc\else\begin{multicols}{2}\fi
		\begin{python3indent4}
class C(object): 
	def __init__(self, *tup):
		self.tup = tup 
	def __eq__(self, other):
		return self.tup == other.tup 
	def __hash__(self): 
		global c; c += 1
		keys = count(self.tup).keys()
		return hash(tuple(keys))?\ifdefined\twocoldoc\else\\\fi?
def tree(n):
	if n == 0: return 0
	return C(tree(n - 1), tree(n - 1))

for h in range(6):
	global c; c = 0 
	tree(h) in {C()}
	print(c)  # 1, 2, 7, 32, 157, 782, ...
		\end{python3indent4}
	\ifdefined\twocoldoc\else\end{multicols}\fi
\end{longlisting}
	Once again, we observe a similar problem as before: the number of operations performed
	is approximately polynomially larger than the size of the data structure.

\subsection{Solution}
	Unlike in the previous comparison example, it is more difficult for
	our \verb|count| algorithm to remedy this problem.
	The solution in this case is to \textit{cache} every object's hash code,
	so that it is not required to be re-computed every time.
	This can be done by either the dictionary itself, by the object itself, or by a proxy
	layer in between.
	This can have its own constant-factor performance implications, but it
	avoids such asymptotic slowdown.

\section{Conclusion}
	It is now clear that constant-factor performance differences can be far more insidious
	than they appear at first sight.
	With the right inputs (or perhaps one should say, with the \textit{wrong} inputs),
	they can be sometimes result in superlinear running times, when only linear
	running times were expected.
	This can have unintended consequences not only for usability, but also for security:
	it can make denial-of-service attacks quite practical.

	Ultimately, this is a lesson about hidden assumptions and implications.
	One reason that such issues often go unnoticed is that hidden assumptions about the
	problem domain frequently preclude them from ever occurring in production.
	Data structures are often implicitly guaranteed to have small depth,
	and short-circuit evaluations can sometimes prevent pathological asymptotic slowdowns
	from occurring in practice.
	Furthermore, where adversarial inputs are involved, attacker-controlled inputs
	often result in a \textit{linear} data structure format, making it difficult
	if not impossible to construct a tree- or graph-like object structure in memory.

	Nevertheless, these issues are of paramount importance when designing general-purpose
	libraries, APIs, or programming language, because implicit regularity assumptions might
	no longer be realistic when there is a desire to leave input data types as unconstrained
	as possible.
	Moreover, even domain-specific applications frequently develop their own utility libraries,
	and those libraries can in turn be used across the codebase, under unexpectedly different
	hidden assumptions that were never explicitly codified or considered when the algorithms
	were first written.

	We therefore suggest that more attention be paid to the full implications of
	traditionally insignificant aspects of code design,
	as something as insignificant as a constant-factor difference in performance can serve as
	a hint of a larger and more fundamental concern that may be at stake.

	\ifdefined\nocustomize
		\bibliographystyle{plain}
	\else
		\nocite{*}
		\bibliographystyle{unsrt}
	\fi
	{\ifdefined\nocustomize\else\small\fi \bibliography{\jobname}}

\appendix

\end{document}